\documentclass[aps,prd,preprintnumbers,superscriptaddress,10pt]{revtex4-2}
\usepackage[utf8]{inputenc}
\usepackage[T1]{fontenc}

\usepackage{amsmath}
\usepackage{amssymb}
\usepackage{booktabs}
\usepackage{graphicx}
\usepackage{xcolor}
\usepackage{lipsum}
\usepackage{lineno}
\usepackage{fancyvrb}
\usepackage[version=4]{mhchem}
\usepackage{chemformula}
\usepackage{isotope}

\usepackage{amssymb}
\usepackage{amsmath}
\usepackage{mathtools}
\usepackage{physics}

\newcommand{\dr}{\mathrm{d}}
\newcommand{\mbf}[1]{\mathbf{#1}}
\newcommand{\unitv}[1]{\hat{\mathbf{#1}}}
\newcommand{\be}{\begin{equation}}
\newcommand{\ee}{\end{equation}}

\usepackage{hyperref}

\DeclareMathOperator{\imag}{Im}

\newcommand{\ignore}[1]{}

\begin{document}

\preprint{HIP-2026-11/TH}

\author{Aula Al-Adulrazzaq}
\email{aula.al-adulrazzaq@helsinki.fi}
\affiliation{Department of Physics, and Helsinki Institute of Physics, University of Helsinki, P.O. Box 64, 00014 Helsinki, Finland}

\author{Paolo Settembri}
\email{paolo.settembri@psi.ch}
\affiliation{PSI Center for Scientific Computing, Theory and Data, Paul Scherrer Institute, 5232 Villigen PSI, Switzerland}
\affiliation{Dipartimento di Scienze Fisiche e Chimiche, Università degli studi dell'Aquila, 67100 L'Aquila, Italy}

\author{Sebastian Sassi}
\affiliation{Department of Physics, and Helsinki Institute of Physics, University of Helsinki, P.O. Box 64, 00014 Helsinki, Finland}

\author{Federico Giannessi}
\affiliation{Dipartimento di Scienze Fisiche e Chimiche, Università degli studi dell'Aquila, 67100 L'Aquila, Italy}

\author{Gianni Profeta}
\affiliation{Dipartimento di Scienze Fisiche e Chimiche, Università degli studi dell'Aquila, 67100 L'Aquila, Italy}
\affiliation{CNR-SPIN L'Aquila, Via Vetoio, 67100 L'Aquila, Italy}

\author{Matti Heikinheimo}
\affiliation{Department of Physics, and Helsinki Institute of Physics, University of Helsinki, P.O. Box 64, 00014 Helsinki, Finland}

\author{Kimmo Tuominen}
\affiliation{Department of Physics, and Helsinki Institute of Physics, University of Helsinki, P.O. Box 64, 00014 Helsinki, Finland}

\title{Anisotropic electron scattering and Migdal effect in semiconductor detectors}

\date{\today}
\begin{abstract}
Cryogenic semiconductor detectors are widely used in dark matter direct detection. Electronic excitations in these materials can be caused either by direct dark matter electron scattering events, or indirectly via dark matter nuclear scattering events. For nuclear scattering of light dark matter, the event rate is enhanced due to an inelastic process known as the Migdal effect. Both the electron recoils and the Migdal effect in semiconductors depend on the dielectric function of the target material via the so-called energy loss function (ELF). In the standard approach found in the literature, the ELF is approximated as isotropic to simplify the calculation of the event rate. 
We introduce a practical method for computing the event rate for a general anisotropic ELF. We find that the daily modulation of the Migdal rate arises solely due to the quadrupole component of the ELF, whereas in electron scattering all spherical harmonic components affect the rate. We apply the formalism to study the daily modulation in silicon and gallium arsenide detectors.
\end{abstract}

\maketitle

\section{Introduction}

Direct detection of sub-GeV dark matter (DM) increasingly relies on target materials capable of resolving electronic excitations at the eV scale or below. Cryogenic semiconductor detectors have emerged as promising platforms \cite{Das_MPLA} due to their sensitivity to single electron-hole pair production and low-energy phononic excitations\cite{Iyer_2021165489,SuperCDMS}. 
Electronic excitations from DM scattering may arise via two distinct processes. Dark matter may interact directly with the target electron, leading to direct DM--electron scattering \cite{Hochberg_PRL, DarkELF, Knapen, Griffin_PRD}. Alternatively, DM scattering off atomic nuclei may lead to electronic excitations. This process can be enhanced for light dark matter due to the Migdal effect \cite{Knapen_Migdal,Berghaus}. Although the Migdal effect has historically eluded observation, very recently, direct observation of the Migdal effect due to neutron scattering in a gaseous target has been achieved \cite{yi-2026}. However, empirical observation of the effect in a solid-state target remains to be realized.

The Migdal effect was originally formulated in the context of isolated atoms \cite{ibe-2018}, where sudden acceleration of the nucleus during a scattering event creates a mismatch between the electronic cloud and the recoiling ion. In atomic systems, electronic wave functions are localized and are naturally expressed in terms of spherical orbitals. Consequently, angular effects are either absent or strongly constrained by atomic symmetry, and the excitation probabilities are often approximately isotropic after averaging over atomic orientations.

In crystalline solids, however, the situation is fundamentally different. Electrons are not localized around isolated nuclei but occupy Bloch states,
$\psi_{n\mathbf{k}}(\mathbf r)=
e^{i\mathbf{k}\cdot\mathbf r}
u_{n\mathbf{k}}(\mathbf r)$,
which depend on the energy quantum number, $n$, and explicitly on crystal momentum $\mathbf{k}$. The electronic response therefore acquires an intrinsic dependence on momentum direction in reciprocal space. Unlike atoms, where the relevant electronic structure can often be described by discrete transitions among localized orbitals, in solids the excitation probability emerges from transitions across extended bands dispersing throughout the Brillouin zone. The anisotropy of the crystal structure, orbital character, and band dispersion can therefore strongly influence the Migdal response.

This distinction becomes particularly important because the Migdal process in solids is commonly formulated through the dielectric response of the material, often expressed through the energy loss function (ELF) of the target material.
The ELF contains information about electronic excitations over a broad range of transferred momenta and energies and naturally incorporates screening and many-body effects. However, in practical calculations, the dielectric response is frequently approximated by its spherical average. Existing implementations, such as the DarkELF framework \cite{DarkELF} and related approaches, employ isotropically averaged responses to reduce computational complexity and simplify the multidimensional phase-space integration.

Although such approximations may be adequate for materials with weak directional dependence, they remove information that could be crucial for understanding the behavior of anisotropic targets. In particular, averaging over momentum directions neglects the dependence on the orientation of the transferred momentum relative to the crystal axes. Since the underlying Bloch wave functions vary throughout reciprocal space, this averaging may hinder strong directional features originating from the electronic structure itself.

The anisotropic response of solids is crucial because the dark matter flux observed in the laboratory is not isotropic. Earth moves through the galactic dark matter halo with a preferred velocity direction, and the orientation of a crystal detector relative to this incoming DM wind changes continuously due to Earth's rotation. As a consequence, anisotropic target responses can translate directly into time-dependent modulation patterns \cite{Boyd_PRD,Heikinheimo_PRD.99.103018}.
These directional signatures may provide several important advantages. First, a daily modulation signal would constitute a powerful discriminator against terrestrial backgrounds, providing evidence for the galactic origin of a signal \cite{Sassi_PRD.104.063037}. Second, directional dependence may encode information about the velocity distribution of dark matter particles beyond what can be extracted from energy spectra alone \cite{Heikinheimo_PRD.109.103010}. Third, anisotropic rates can reveal material-specific fingerprints inaccessible in isotropic calculations. Different crystal orientations, symmetries, or electronic structures may generate distinct modulation amplitudes and phases, opening a new pathway toward dark matter characterization.

Moreover, anisotropic Migdal responses could produce signatures beyond simple overall rate variations. Sharp directional structures in reciprocal space may enhance particular momentum-transfer channels, induce orientation-dependent thresholds, or generate energy-dependent modulation patterns. Such effects may substantially alter the expected event spectra and affect the inferred sensitivity of future experiments. In this sense, anisotropy is not merely a correction to the total rate but may constitute an entirely new source of experimental information.

Schematically, a generic DM scattering process that encompasses both direct electron scattering and the Migdal effect can be described in terms of a momentum $\mbf{q}$ and an associated energy $\omega_q$ deposited in the target, releasing an excitation with energy $\omega$. The scattering rate can then be expressed in the form \cite{Trickle_PRD.101.055004,Kahn_2022}
\begin{equation}
    \label{eq:general-dm-rate}
    \frac{\dr R}{\dr \omega} \sim \int\dr^3\mbf{q}\,\frac{F(q)}{q}\int\dr\omega_q\,T(\mbf{q},\omega_q,\omega)\hat{f}(s + \mbf v_\text{lab}\cdot\unitv{q},\unitv{q}),
\end{equation}
where $T(\mbf{q},\omega_q,\omega)$ represents a generalized target response, encoding the details of the microscopic scattering process, and $\hat{f}$ is a Radon transform of the DM velocity distribution, with $s = q^2/2m_\chi + \omega_q/q$ and $\mbf{v}_\text{lab}$ denoting the laboratory velocity in the galactic frame.

The challenge associated with evaluating the scattering rate is the combination of the multi-dimensional integrals over the momentum and energy transfer, even when the Radon transform of the velocity distribution has a known closed form expression. This is in conjunction with the fact that, for any meaningful analysis, the event rate typically needs to be evaluated at multiple energies and multiple points in time for multiple DM models.

In this work, we describe a framework for evaluating the event rate in the form of \eqref{eq:general-dm-rate} while retaining the directional dependence of the target response. Our analysis employs a spherical harmonic expansion of the generalized target response. Applied to DM--electron scattering and the Migdal effect, this allows us to draw a connection between the multipoles of the ELF and the resulting daily modulation. In particular, we demonstrate that in case of the Migdal effect, the daily modulation is, to a good approximation, sensitive only to the quadrupole component of the ELF. In addition, we have computed the resulting daily modulation signals in silicon (Si) and gallium arsenide (GaAs) detectors in both electron scattering and the Migdal effect.

Furthermore, we note that efficient computational methods are required for large-scale analyses of daily modulation due to detector anisotropies. The spherical harmonic expansion method employed in the framework described in this work is straightforward to adapt to fast event rate computation algorithms, which rely on similar harmonic expansions, such as that presented in Ref. \cite{Sassi_jzmc-r2k6}.

The paper is organized as follows: In Sec.~\ref{section:theory} we discuss the formalism for evaluation of the event rates. We start with a description of the computational details of the dielectric function which is the relevant input of condensed matter for the description of the target structure. We then proceed to formulate the event rate for direct electron scatterings, and then consider our main case of interest, the Migdal effect in a solid state target. In both cases our results are concise formulas that allow controlled evaluation of the daily modulation of the event rates taking into account anisotropies from the target structure and the galactic DM velocity distribution. In Sec.~\ref{section:results} we illustrate the application of these formulas, and in Sec.~\ref{section:checkout} we present our conclusions and outlook for future work.

\section{Event rate calculation}
\label{section:theory}

\subsection{General formalism}

In this section, we present the general formulae for direct detection event rates. Although the electron recoil rate and the Migdal rate arise from two distinct processes, their formulae can be described in a process-agnostic way due to their common origin in quantum mechanical scattering theory \cite{Trickle_PRD.101.055004,Kahn_2022}. We will thus describe DM scattering off a generic target, which we then specialize to both processes. Specifically, we consider a process in which a DM particle transfers momentum $\mbf{q}$ and energy $\omega_q$ to a target, generating an excitation with momentum $\mbf{k}$ and energy $\omega$, which corresponds to the measured energy, accounting for potential loss of energy $\omega_q - \omega$ to some other channel. We give the  differential rate for such a process in the form
\begin{equation}
    \frac{\dr R}{\dr \omega} \sim \int \dr^3\mbf{q}\,F(q^2)\int\dr\omega_q\int\dr^3\mbf{k}\,\mathcal{W}(\mbf{k},\omega)S(\mbf{q} - \mbf{k},\omega_q - \omega)g(\mbf{k},\mbf{q})\int\dr^3\mbf{v}\,f(\mbf{v})\delta\left(\mbf{v}\cdot\mbf{q}-\frac{q^2}{2m_\chi}-\omega_q\right).
    \label{eq:gen-rate-full}
\end{equation}
Here, $\mathcal{W}(\mbf{k},\omega)$ represents the response of the target material to the energy deposition $\omega$, while $S(\mbf{q} - \mbf{k},\omega_q - \omega)$ represents the response to the lost energy $\omega_q - \omega$. Note that for composite targets consisting of multiple target types (e.g., elements or isotopes) the total event rate is given by a sum of the rates for the individual target types weighted by their relative abundances in the composite.

The function $F(q^2)$ is a generic form factor that captures deviations from pointlike interactions of the DM with the target. The rates for electron and Migdal scattering are typically suppressed for momentum transfers above $\mathcal{O}(10\text{ keV})$. Therefore, we may treat the target as point-like even for nuclear scattering. Furthermore, assuming standard spin-independent interactions, it follows that the form factor reduces to the dark matter form factor $F_\text{DM}(q^2) = (q_0^2 + M^2)/(q^2 + M^2)$, where $M$ is the mass of a mediator. In this work we focus exclusively on the heavy mediator limit $F_{\text{DM}}(q^2)\approx 1$, which for low momentum transfers applies down to relatively low mediator masses.

Concretely, in this work $\omega$ is the energy of an outgoing electron, and $\mathcal{W}(\mbf{k},\omega)$ is the energy loss function (ELF) of the target material, defined in terms of the dielectric function $\varepsilon(\mbf{k}, \omega)$ as
\begin{equation}
    \mathcal{W}(\mbf{k}, \omega) = -\imag\left(\frac{1}{\varepsilon(\mbf{k}, \omega)}\right).
    \label{eq:ELFeq}
\end{equation}
In case of the Migdal effect, $S(\mbf{q} - \mbf{k},\omega_q - \omega)$ corresponds to the dynamic structure factor, which captures the deposition of energy into vibrational modes of the crystal from the nuclear recoil, and $g(\mbf{k},\mbf{q})$ is a factor derived from the coupling of the electron cloud with the nucleus. In the case of electron recoils, the DM deposits its energy directly to the electron without an intermediate nuclear recoil, and we can set $S(\mbf{q} - \mbf{k},\omega_{q} - \omega)g(\mbf{k},\mbf{q}) = \delta(\omega_q - \omega)\delta(\mbf{q} - \mbf{k})$.

The principal idea behind this schematic description is that we can divide the process into a generalized target response function
\begin{equation}
    T(\mbf{q},\omega_q,\omega) = \int\dr^3\mbf{k}\,\mathcal{W}(\mbf{k},\omega)S(\mbf{q} - \mbf{k},\omega_{q} - \omega)g(\mbf{k},\mbf{q}),
    \label{eq:gen-response}
\end{equation}
which depends only on the target material and its response to the scattering, and the term
\begin{equation}
    \frac{1}{q}\hat{f}(s,\unitv{q}) = \int\dr^3\mbf{v}\,f(\mbf{v})\delta\left(\mbf{v}\cdot\mbf{q}-\frac{q^2}{2m_\chi}-\omega_{q}\right),
    \label{eq:gen-radon}
\end{equation}
where $s = q/2m_\chi + \omega_{q}/q$, and $\hat{f}(s,\unitv{q})$ is the Radon transform of the velocity distribution, which depends only on the scattering kinematics and the local distribution of dark matter. This nicely encapsulates the parts of the event rate integral that depend on the target, and parts that depend on DM kinematics.

Typically, the DM velocity distribution is defined in the galactic frame, whereas the event rate \eqref{eq:gen-rate-full} is evaluated in an Earth-based laboratory frame, which is boosted relative to the galactic frame by the laboratory velocity in the galactic frame $\mbf v_\text{lab}(t)$. Under such a boost, given the galactic frame velocity distribution $f(\mbf{v})$, the Radon transform transforms as $\hat{f}(s,\unitv{q}) \rightarrow \hat{f}(s + \mbf v_\text{lab}\cdot\unitv{q},\unitv{q})$ \cite{Gondolo_PRD.66.103513}. The general event rate formula in the laboratory frame can then be written as
\begin{equation}
    \frac{\dr R}{\dr \omega} \sim \int\dr^3\mbf{q}\,\frac{F(q^2)}{q}\int\dr\omega_q\,T(\mbf{q},\omega_q,\omega)\hat{f}(s + \mbf v_\text{lab}\cdot\unitv{q},\unitv{q}).
    \label{eq:gen-rate}
\end{equation}

For the discussion of the target anisotropy and its impact on the event rate, it is advantageous to express the formula \eqref{eq:gen-rate} in a spherical harmonic basis. Since all the functions involved are real, it is natural to employ the real spherical harmonics $Y_{\ell m}(\unitv{q})$, related to the complex spherical harmonics $Y_\ell^m(\unitv{q})$ via
\begin{equation}
    Y_{\ell m} = 
        \begin{cases}
            \frac{i}{\sqrt{2}}(Y_\ell^m - (-1)^mY_\ell^{-m}) & m < 0, \\
            Y_\ell^0 & m = 0, \\
            \frac{1}{\sqrt{2}}(Y_\ell^{-m} + (-1)^mY_\ell^m) & m > 0.
        \end{cases}
    \label{eq:real-sh}
\end{equation}
In this work, we use unit normalized spherical harmonics which include the Condon--Shortley phase. We now write the functions in the integrand of equation \eqref{eq:gen-rate} using their spherical harmonic expansions
\begin{equation}
    T(\mbf{q},\omega_q,\omega) = \sum_{\ell m}T_{\ell m}(q,\omega_q,\omega)Y_{\ell m}(\unitv{q})
    \label{eq:gen-response-harmonics}
\end{equation}
and
\begin{equation}
    \hat{f}(s + \mbf v_\text{lab}\cdot\unitv{q},\unitv{q}) = \sum_{\ell m}\hat{f}_{\ell m}(s,\mbf v_\text{lab})Y_{\ell m}(\unitv{q}).
    \label{eq:gen-radon-harmonics}
\end{equation}

Because the target response is naturally defined in the lab frame, whereas the velocity distribution is naturally defined in the galactic frame, evaluation of the event rate always involves a time-dependent rotation to a common coordinate system. For the spherical harmonic coefficients, this means rotation of the coefficients by the real analog of Wigner's $D$-matrix, $\mathcal{R}^{(\ell)}_{mm'}(t)$, related to the coefficients $D^{(\ell)}_{mm'}$ by the linear transformation \eqref{eq:real-sh} that defines the real spherical harmonics. It is useful to define the coefficients $\hat{f}_{\ell m}$ in a frame whose $z$-axis is aligned with $\mbf v_\text{lab}$, since then $\hat{f}_{\ell m}$ only depend on the magnitude of $\mbf v_\text{lab}$. The event rate at an arbitrary moment in time is then given by  
\begin{equation}
    \frac{\dr R}{\dr \omega} \sim \int\dr q\,qF(q^2)\int\dr\omega_q\,\sum_{\ell mm'}T_{\ell m}(q,\omega_q,\omega)\mathcal{R}^{(\ell)}_{mm'}(t)\hat{f}_{\ell m'}(s,v_\text{lab}).
    \label{eq:gen-rate-harmonic}
\end{equation}
This expression pairs the multipoles of the material response with the multipoles of the Radon transform, which arise from the anisotropy of the scattering kinematics and from the local dark matter velocity distribution. As such, it also makes clear that the profile of the daily modulation signal is primarily determined by which multipoles of the material response are relevant and which are suppressed.

Although the above discussion is fully general and applies to any velocity distribution, in this work, we work in the Standard Halo Model (SHM), which has the velocity distribution
\begin{equation}
    f_\text{SHM}(\mbf{v}) = \frac{1}{N_\text{esc}\sqrt{(\pi v_0)^3}}e^{-\mbf{v}^2/v_0^2}\Theta(v_\text{esc} - |\mbf{v}|) ,
    \label{eq:fSHM}
\end{equation}
where the normalization constant $N_\text{esc}$ is given by
\begin{equation}
N_{\rm esc} = \erf\left(\frac{v_{\rm esc}}{v_0}\right) - \frac{2}{\sqrt{\pi}}\frac{v_{\rm esc}}{v_0}\exp\left(-\frac{v_{\rm esc}^2}{v_0^2}\right),
\end{equation}
and $v_{\rm esc}$ is the galactic escape velocity. The Radon transform of the SHM in the laboratory frame has a closed form solution given by
\begin{equation}
    \hat{f}_\text{SHM}(s +\mbf v_\text{lab}\cdot\hat{\mbf q}, \hat{\mbf q}) = \frac{1}{N_{\rm esc} \sqrt\pi v_0} \left( \exp\left[-\frac{(s+\mbf v_\text{lab}\cdot\hat{\mbf q})^2}{v_0^2}\right] - \exp\left[-\frac{v_{\rm esc}^2}{v_0^2}\right] \right) \Theta(v_{\rm esc}-|s+\mbf v_\text{lab}\cdot\hat{\mbf q}|),
\end{equation}
where $\mbf v_\text{lab}(t)$ is the velocity of the laboratory in the galactic frame.

When it comes to the evaluation of the spherical harmonic coefficients
\begin{equation}
\hat{f}_{\ell m}^\text{SHM}(w,v_\text{lab}) = \int\dr\Omega\,\hat{f}_\text{SHM}(s + \mbf v_\text{lab}\cdot\unitv{q},\unitv{q})Y_{\ell m}(\unitv{q}),
\end{equation}
given the step function $\Theta(v_{\rm esc}-|s+\mbf v_\text{lab}\cdot\hat{\mbf q}|)$ that appears in the Radon transform, the limits of these integrals generally become nontrivial when restricted to the support of the Radon transform. Therefore, it makes sense to evaluate the integrals in a coordinate system, where the $z$-axis is along $\mbf v_\text{lab}(t)$. In this case, the condition of the step function reduces to
\begin{equation}
    |v_\text{lab}(t)\cos\theta + s| \leq v_\text{esc}.
\end{equation}
The azimuthal integral in these coordinates is therefore unaffected and goes from 0 to $2\pi$, while the polar integral is limited to
\begin{equation}
    -\frac{v_\text{esc} + s}{v_\text{lab}(t)} \leq \cos\theta \leq \frac{v_\text{esc} - s}{v_\text{lab}(t)}.
\end{equation}
The expression for the spherical harmonic coefficients is then reduced to
\begin{equation}
\hat{f}_{\ell m}^\text{SHM}(s,v_\text{lab}) = 2\pi\delta_{m0}\int_{z_\text{min}}^{z_\text{max}}\dr z\,\hat{f}_\text{SHM}(s + v_\text{lab}z)P_\ell(z),
\end{equation}
where $P_l(z)$ are Legendre polynomials and
\begin{equation}
    z_\text{min}=\max\left\{-1,-\frac{v_\text{esc}+s}{v_\text{lab}(t)}\right\}, \qquad
    z_\text{max}=\min\left\{1,\frac{v_\text{esc}-s}{v_\text{lab}(t)}\right\}.
\end{equation}
Given that these are integrals of a Gaussian plus a constant function multiplied by a polynomial, we note that it is, in principle, possible to write down a general closed form expression for arbitrary $\ell$ using known expressions for Legendre polynomials. However, we will not write down such an expression here, since there are more practical methods of generating these coefficients. We do, however, give expressions for the $\ell = 0, 2$ terms in the Appendix since these serve a special role in the Migdal effect in the soft limit.

In the following subsections, we will apply the above formalism to the creation of electronic excitations in two specific cases: first, to the scattering of DM particles directly off electrons in the target material, and second, to the Migdal effect; that is, creation of electronic excitations as a consequence of DM scattering off nuclei. We will demonstrate how the generalized target response $T(\mbf{q},\omega_q,\omega)$ in these two scenarios depends on the ELF in different ways, leading to distinct but analogous daily modulation behavior.

\subsection{Dielectric function calculations}
\label{subsection:DFT}

Although common in the literature, the use of the free electron gas model \cite{FEG} or fits from experimental data, such as those using Mermin oscillators \cite{Mermin}, to model the ELF do not include any information on the angular dependency of the material response.
However, density functional theory \cite{HohenbergKohn,KohnSham,TDDFT} offers a common and reliable tool that enables first-principles calculations of the lattice and electronic structure of a material, as well as its dielectric function \cite{Adler,Wiser,Hybertsen}.
Therefore, given our need for the anisotropic dielectric function of a material to compute the ELF via Eq. \eqref{eq:ELFeq}, we need to use DFT calculations for its evaluation. To accomplish this, we use the publicly available GPAW code for DFT calculations \cite{GPAW,ASE}, which is widely used in the DM community for accurate evaluation of electronic effects.

An accurate computation of the event rate requires the evaluation of the anisotropic dielectric function over a wide range of electronic momenta. In crystalline systems, reciprocal space is fundamentally partitioned into Brillouin zones (BZs); thus, capturing these large momentum transfers necessitates covering multiple BZs.

In practice, a radial cutoff must be imposed in reciprocal space, which means the dielectric function is evaluated for all points up to a maximum magnitude of the momentum, $q_{\text{cut}}$. However, the computational cost and memory required for the calculation increase cubically with $q_{\text{cut}}$, making well-converged calculations difficult to achieve for large momentum cut-offs.

To better tackle these critical points, we implemented the use of crystal symmetries to reduce the number of required calculations without any loss of accuracy.
Specifically, in a periodic crystal, the dielectric response is described in terms of the microscopic dielectric matrix, $\varepsilon_{\mathbf{G}\mathbf{G}'}(\mathbf{k}, \omega)$. The momentum transfer is decomposed into the vector $\mathbf{k}$, which is sampled from a discretized grid within the first BZ, and the reciprocal lattice vectors $\mathbf{G}$ and $\mathbf{G'}$, which are defined by the unit cell of the system. This formulation allows us to capture local field effects arising from the lattice periodicity.

Using crystal symmetries, the $\mathbf{k}$-point sampling can be reduced from the BZ to the so-called irreducible Brillouin zone (IBZ), significantly lowering the computational burden. We now describe such crystal symmetries and show how the microscopic dielectric function transforms under them.

Crystal symmetries are generally expressible as an affine transformation $\mathbf{r}'=\hat{M}\mathbf{r}+\boldsymbol{t}$, consisting of a linear transformation given by the matrix $\hat{M}$, which describes a rotation or a reflection, and a shift given by a fractional lattice vector $\boldsymbol{t}$, which is present only in the case of non-symmorphic transformations (such as screw axes and glide reflections). Given the real-space definition of the dielectric function and these symmetry operations, it is straightforward to derive how the symmetry operation will act on the dielectric function in reciprocal space \cite{PhDThesis_Settembri},
\begin{equation}
\varepsilon_{\mathbf{G},\mathbf{G}'}(\mathbf{k},\omega)=e^{-i2\pi(\mathbf{G}-\mathbf{G}')\cdot\boldsymbol{t}} \, \varepsilon_{\hat{U}\mathbf{G},\hat{U}\mathbf{G}'}(\hat{U}\mathbf{k},\omega),
\label{eq:epsrotfrac}
\end{equation}
where $\hat{U}$ is an integer matrix (containing only $-1$, 0, and 1) that represents the point-group operation $\hat{M}$ in reciprocal space. For clarity, all spatial and momentum vectors are expressed in fractional coordinates.

To exploit the relations reported in Eq. \eqref{eq:epsrotfrac}, we modified the GPAW source code to explicitly output the microscopic dielectric matrices. We then developed a dedicated post-processing script to reconstruct the full BZ response by applying the appropriate symmetry operations on the irreducible grid.

Let $\mathbf{k}$ and $\mathbf{k}_{\text{irr}}$ be the momentum grid points inside the full BZ and the irreducible BZ, respectively. The geometric mapping adopted by the GPAW code linking the two points can be written as
\begin{equation}
\begin{cases}
       \hat{U}_{s} \mathbf{k} = \mathbf{k}_{\text{irr}} + \mathbf{N}, \\
       \mathbf{k} = \hat{U}_{r} \mathbf{k}_{\text{irr}} + \hat{U}_{r}\mathbf{N} = \hat{U}_{r} \mathbf{k}_{\text{irr}} + \mathbf{N}',
\end{cases}
\label{eq:mapping}
\end{equation}
where $\hat{U}_{s}$ indicates the $s$-th symmetry operation matrix of the system, while $\hat{U}_{r}=\hat{U}_{s}^{-1}$ is its inverse and the $r$-th symmetry operation. The reciprocal lattice vector $\mathbf{N}$, which in fractional units is a vector of integers, is used in case $\hat{U}_s\mathbf{k}$ falls outside of the first Brillouin zone, shifting the vector back inside. We also define $\mathbf{N}'\equiv\hat{U}_r\mathbf{N}$. Combining the symmetry rules of the dielectric function from Eq.~\eqref{eq:epsrotfrac} with the mapping between points in the irreducible BZ and the BZ of Eq. \eqref{eq:mapping}, we obtain the following relations
\begin{equation}
    \begin{cases}        
           \varepsilon_{\mathbf{G},\mathbf{G}'}(\mathbf{k},\omega)=e^{-i2\pi(\mathbf{G}-\mathbf{G}')\cdot\boldsymbol{t}_s} \ \varepsilon_{\hat{U}_s\mathbf{G}+\mathbf{N},\hat{U}_s\mathbf{G}'+\mathbf{N}}(\mathbf{k}_{\text{irr}},\omega), \\
           \varepsilon_{\mathbf{G},\mathbf{G}'}(\mathbf{k}_{\text{irr}},\omega)=e^{-i2\pi(\mathbf{G}-\mathbf{G}')\cdot\boldsymbol{t}_r}    \ \varepsilon_{\hat{U}_{r}\mathbf{G}-\mathbf{N}',\hat{U}_{r}\mathbf{G}'-\mathbf{N}'}(\mathbf{k},\omega).
    \end{cases}
    \label{eq:epsrot}
 \end{equation}
Eq. \eqref{eq:epsrot} allows us to obtain the dielectric function on a full $\mathbf{k}$-point grid in reciprocal space by only evaluating it on the irreducible BZ points, $\mathbf{k}_{\text{irr}}$, reducing the computational burden without any accuracy loss.

To handle local field effects (LFEs) more efficiently, we configured the code to directly output the LFE-included dielectric function, defined as
\begin{equation}
    \varepsilon( \mbf k + \mbf G, \omega) \equiv \varepsilon^{\text{LFE}}_{\mathbf{G},\mathbf{G}}(\mathbf{k},\omega)=\frac{1}{[\varepsilon^{-1}(\mathbf{k},\omega)]_{\mathbf{G},\mathbf{G}}} .
    \label{eq:LFEs}
\end{equation}
Since we are considering diagonal elements, the phase factor present in Eq. \eqref{eq:epsrot} will vanish also for non-symmorphic operations, symplifying the reconstruction of the full BZ with some modifications made by us that improve the dielectric function description and reduce the computational cost.

\begin{figure}
    \centering
    \includegraphics[width=0.99\linewidth]{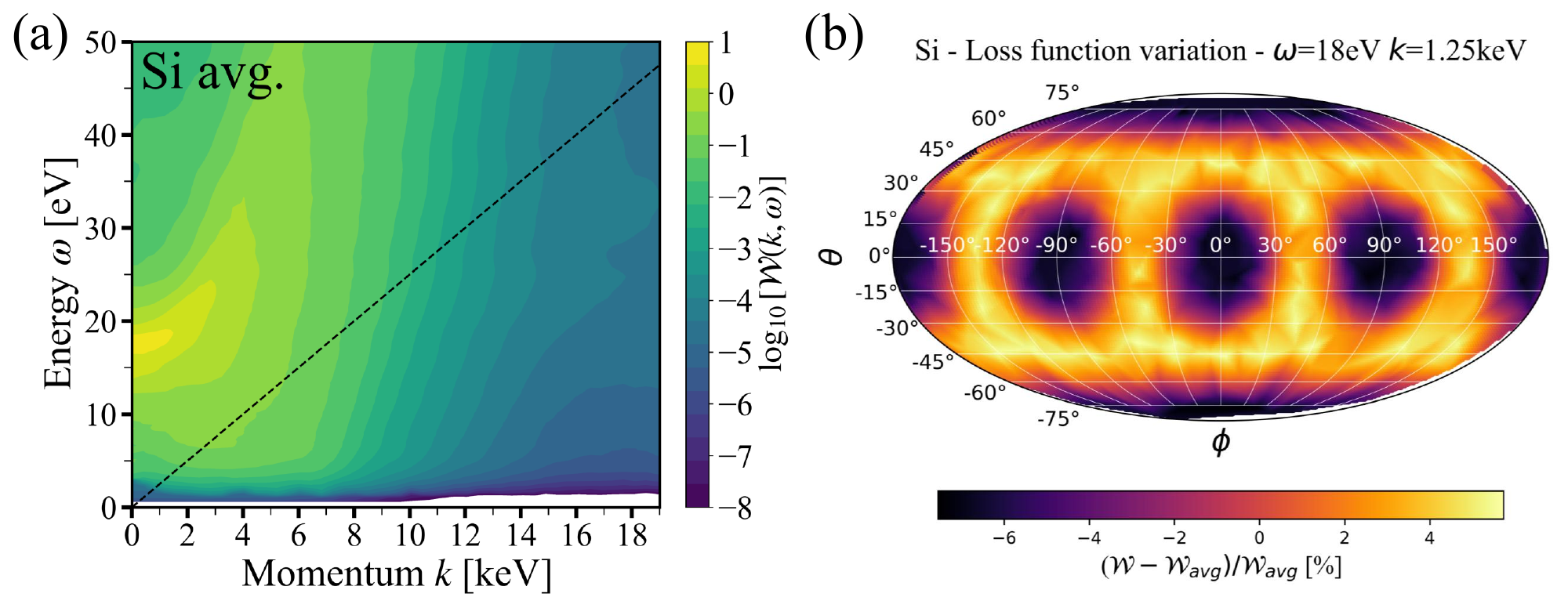}
    \caption{(a) Spherically averaged electronic loss function of Si. The black dashed line indicates the upper kinematic boundary for DM-e$^{-}$ scattering. (b) Mollweide projection of the ELF fractional variation with respect to the spherical average, evaluated at $\omega$=18 eV and $k$=1.25 keV.}
    \label{fig:Si-DF}
\end{figure}

Although we compute the scattering rates using the ELF evaluated on a regular grid, we adopted an interpolation scheme to better visualize its spherical average and angular modulation. Specifically, we uniformly sample points over spheres in momentum space using the Fibonacci lattice method \cite{gonzalez-2009}, perform a linear interpolation, and then average over spheres of constant radius. Fig. \ref{fig:Si-DF} (a) shows the spherically averaged ELF of silicon (Si), computed using the DFT parameters detailed in Tab. \ref{tab:dft_parameters}. Fig. \ref{fig:Si-DF} (b) presents a Mollweide projection of the relative deviation of the ELF from its average value for a fixed energy and momentum value.

\subsection{Electron recoils}
\label{subsection:electrons}

The differential event rate per unit target mass for DM--electron scattering can be expressed as \cite{Kahn_2022}
\begin{equation}
    \label{eq:electron-rate-integral}
    \frac{\dr R_{\rm e}(t)}{\dr \omega}  = \frac{1}{\rho_T} \frac{\rho_\chi}{m_\chi} \frac{\bar{\sigma}_e}{\mu_{\chi e}^2} \frac{1}{2\alpha} \int \frac{\dr^3 \mbf{q}}{(2\pi)^3} F_{DM}^2(q, \alpha m_e) q^2 \mathcal{W}(\mbf{q}, \omega) \int \dr^3 \mbf{v} \delta\left(\omega-\mbf{q} \cdot \mbf{v} + \frac{q^2}{2m_\chi}\right) f(\mbf{v}+\mbf v_\text{lab}(t)).
\end{equation}
Here, $\rho_T$ denotes the mass density of the target material, $\rho_\chi$ is the local DM energy density, $\alpha$ is the fine structure constant, and $\mu_{\chi e}=m_\chi m_e/(m_\chi+m_e)$ is the reduced mass of the DM--electron system, where $m_\chi$ is the mass of the dark matter particle. The DM--electron interaction is parametrized by the reference cross section $\bar\sigma_e$, and the form factor $F_{\rm DM}$, which we set to $F_{\rm DM}^2 \simeq 1$ corresponding to the heavy mediator limit.

In the context of the general event rate formula \eqref{eq:gen-rate}, equation \eqref{eq:electron-rate-integral} corresponds to setting the generalized target response $T(\mbf{q},\omega_q,\omega) = \mathcal{W}(\mbf{q},\omega)\delta(\omega_q - \omega)$, that is
\begin{equation}
    \label{eq:electron-rate-short}
    \frac{\dr R_{\rm e}(t)}{\dr \omega} =  \frac{\rho_\chi \bar\sigma_e}{16 \pi^3\alpha\rho_T m_\chi \mu_{\chi e}^2}   \int \dr^3 \mbf q \, q \, \mathcal{W}(\mbf q, \omega) \hat{f}(s + \mbf v_\text{lab}\cdot\unitv{q},\hat{\mbf q}),
\end{equation}
or in terms of the spherical harmonic expansion \eqref{eq:gen-rate-harmonic},
\begin{equation}
    \label{eq:electron-rate-short-harmonic}
    \frac{\dr R_{\rm e}(t)}{\dr \omega} =  \frac{\rho_\chi \bar\sigma_e}{16 \pi^3\alpha\rho_T m_\chi \mu_{\chi e}^2}   \int \dr q \, q^3 \, \sum_{lmm'}\mathcal{W}_{lm}(q, \omega)\mathcal{R}^{(\ell)}_{mm'}(t)\hat{f}_{\ell m'}(s,v_\text{lab}).
\end{equation}
We note that the special case of an isotropic ELF with $\mathcal{W}(\mbf{q},\omega) = \mathcal{W}(q,\omega) = \mathcal{W}_{00}(q,\omega)/\sqrt{4\pi}$, is obtained by neglecting all terms with $\ell > 0$ in the sum, such that
\begin{equation}
    \label{eq:electron-rate-short-iso}
    \frac{\dr R_{\rm e}^\text{iso}(t)}{\dr \omega} =  \frac{\rho_\chi \bar\sigma_e }{8 \pi^{5/2}\alpha\rho_T m_\chi \mu_{\chi e}^2}   \int \dr q \, q^3 \, \mathcal{W}(q, \omega)\hat{f}_{00}(s,v_\text{lab}).
\end{equation}
Given that there is generally no closed form expression for the ELF, the radial integrals in \eqref{eq:electron-rate-short-harmonic} and \eqref{eq:electron-rate-short-iso} must in general be evaluated numerically. In the special case of Eq.\eqref{eq:electron-rate-short-iso}, the isotropic ELF projects out the isotropic part of the velocity distribution. Conversely, only materials with an anisotropic ELF would be sensitive to anisotropies of the DM velocity distribution when considering DM-electron scattering.

\subsection{Migdal effect}
\label{subsection:migdal}

In case of the Migdal effect, the expression for the differential event rate per unit target mass for electron excitations is given by \cite{Berghaus} 
\begin{equation}
\begin{split}
     \frac{\dr R_{\rm M}}{\dr\omega} &= \frac{\rho_\chi}{m_\chi}\frac{ \bar\sigma_{ n}}{\mu^2_{\chi n}} \frac{A^2}{m_N^3} \frac{8\pi^2\alpha}{\omega^4} 
     \int \dr^3 \mbf v f(\mbf v + \mbf v_\text{lab}(t)) \int \frac{\dr^3 \mbf q}{(2\pi)^3} F_{DM}^2(q, m_\chi v_0) \int_{1BZ} \frac{\dr^3\mbf k_e}{(2\pi)^3} \sum_{\mbf K} Z^2(|\mbf k_e + \mbf K|)\\
    &\quad\times \frac{(\mbf q\cdot(\mbf k_e+\mbf K))^2}{|\mbf k_e+\mbf K|^2} \mathrm{Im}[-\varepsilon^{-1}_{\mbf K \mbf K}(\mbf k_e,\omega)] S\left(\mbf q-\mbf k_e-\mbf K, \mbf q\cdot \mbf v - \frac{q^2}{2m_\chi}-\omega\right) ,
\end{split}
    \label{eq:migdal-rate}
\end{equation}
where $\mu_{\chi n}$ is the reduced mass of the DM--nucleon system, $A$ is the atomic mass number, and $m_N$ is the mass of the nucleus. Here, $Z(k)$ denotes the momentum-dependent effective ion charge, which accounts for the net electrostatic contribution of the bare nucleus partially screened by the core electrons. The integral over the electron momentum $\mbf{k}_e$ is taken over the first Brillouin zone, and the sum over the reciprocal lattice vectors $\mbf{K}$ then covers the entire momentum space.

We wish to write the rate \eqref{eq:migdal-rate} in the form of \eqref{eq:gen-rate-full} to apply the formulae from the beginning of this section. This can be accomplished by inserting the integral
\begin{equation}
    \int\dr\omega_q\,\delta\left(\omega_q - \mbf{v}\cdot\mbf{q} + \frac{q^2}{2m_\chi}\right) = 1.
\end{equation}
into \eqref{eq:migdal-rate}. This allows us to identify the generalized target response $T(\mbf{q},\omega_q,\omega)$ of equation \eqref{eq:gen-rate} as
\begin{equation}
    T(\mbf{q},\omega_q,\omega) = \int_{1BZ} \dr^3\mbf{k}_e \sum_{\mbf{K}} Z^2(|\mbf{k}_e + \mbf{K}|)\frac{(\mbf{q}\cdot(\mbf{k}_e + \mbf{K}))^2}{|\mbf{k}_e + \mbf{K}|^2}\mathcal{W}(\mbf{k}_e + \mbf{K},\omega) S(\mbf{q} - \mbf{k}_e-\mbf{K}, \omega_q - \omega),
    \label{eq:gen-response-migdal}
\end{equation}
where we have identified the diagonal ELF $\mathcal{W}(\mbf{k}_e + \mbf{K},\omega) = \imag[-\varepsilon^{-1}_{\mbf{K}\mbf{K}}(\mbf{k}_e,\omega)]$. From now on, we also combine the integral of $\mbf{k}_e$ over the first Brillouin zone with the sum over the reciprocal lattice vectors $\mbf{K}$ into a single integral over all momenta $\mbf{k} = \mbf{k}_e + \mbf{K}$. The event rate then takes the form
\begin{equation}
    \frac{\dr R_\text{M}}{\dr\omega} = \frac{\rho_\chi}{m_\chi}\frac{ \bar\sigma_{ n}}{\mu^2_{\chi n}} \frac{A^2}{m_N^3} \frac{\alpha}{\omega^4}\frac{1}{8\pi^4}\int\dr^3\mbf{q}\, \dr\omega_q\, \frac{1}{q}T(\mbf{q},\omega_q,\omega)\hat{f}(s + \mbf v_\text{lab}\cdot\unitv{q},\unitv{q}),
\end{equation}
where $s = \omega_q/q + q/2m_\chi$, and we have assumed the heavy mediator limit $F_{DM}^2(q,m_\chi v_0)\approx 1$.

The advantage of the general formalism is evident here, since now all the complexity of electronic excitations and lattice vibrations has been banished to the generalized target response \eqref{eq:gen-response-migdal}. Even in the case where no approximations can be made to simplify the expression, as long as it is possible to tabulate the values of $T(\mbf{q},\omega_q,\omega)$ for a given material in some useful way, computation of the Migdal rate is not substantially more difficult or too different from computation of the electron rate.

The dynamic structure factor of the lattice is present in equation \eqref{eq:gen-response-migdal}, because part of the transferred momentum $\mbf{q}$ and energy $\omega_q$ is transferred to the atomic lattice as vibrations. This function is formed out of the lattice wave functions, and is typically written using approximate formulae \cite{Berghaus}. For nuclear recoil energies $E_r$ significantly higher than the average phonon energy, the nuclei may be treated as free. In this approximation $S(\mbf{q} - \mbf{k}, \omega_q - \omega)$ reduces to a delta-function,
\begin{equation}
    \label{eq:free-ion}
    S\left(\mbf{q} - \mbf{k}, \omega_q - \omega\right) \simeq 
    \delta\left(\omega_q - \omega - \frac{(\mbf{q} - \mbf{k})^2}{2m_N}\right).
\end{equation}
This expression may, in general, be used to reduce the dimensionality of the $\mbf{k}$ integral in equation \eqref{eq:gen-response-migdal}. However, it is typically also feasible to employ the soft approximation, and assume that the electron momentum $\mbf{k}$ is small relative to the momentum transfer $\mbf{q}$, in which case we may take $\mbf{q} - \mbf{k} \approx \mbf{q}$ in the dynamic structure factor. Then equation \eqref{eq:free-ion} and the integral over $\omega_q$ amount to setting $\omega_q \approx \omega + q^2/2m_N$ such that the generalized target response reduces to
\begin{equation}
    T(\mbf{q},\omega_q,\omega) = \delta\left(\omega_q - \omega - \frac{q^2}{2m_N}\right)\int \dr^3\mbf{k}\, Z^2(k)\frac{(\mbf{q}\cdot\mbf{k})^2}{k^2}\mathcal{W}(\mbf{k},\omega)
    \label{eq:gen-response-migdal-soft}
\end{equation}
Expanded, the differential event rate now reads 
\begin{equation}\label{eq:6D}
    \frac{\dr R_{\rm M}}{\dr\omega} = \frac{A^2 \alpha \bar\sigma_n \rho_\chi}{8\pi^4  \mu_{\chi n}^2 m_N^3 m_\chi \omega^4} \int \dr^3 \mbf{q} \int \dr^3 \mbf{k} \, Z^2(k)  \mathcal{W}(\mbf{k}, \omega)  \frac{(\mbf{q}\cdot \mbf{k})^2}{q k^2} \hat{f}(s + \mbf v_\text{lab}\cdot\unitv{q},\unitv{q}),
\end{equation}
where, given the delta-function of equation \eqref{eq:gen-response-migdal-soft}, we can write
\begin{equation}
    s = \frac{q}{2\mu_{\chi N}} + \frac{\omega}{q}.
    \label{eq:soft-limit}
\end{equation}

To briefly justify use of the soft limit approximation, we first note that it relates to the general kinematic condition
\begin{equation}
    \mbf{v}\cdot\mbf{q} - \frac{q^2}{2m_\chi} - \frac{(\mbf{q} - \mbf{k})^2}{2m_N} - \omega 
    = \mbf{v}\cdot\mbf{q} - \frac{q^2}{2\mu_{\chi N}} + \frac{\mbf{q}\cdot\mbf{k}}{m_N} - \frac{k^2}{2m_N} - \omega = 0.
\end{equation}
This condition, resulting from energy conservation, determines the kinematic region the event rate integral probes. The relevant terms for the soft limit are the three middle terms. It is clear that the magnitude of the $\mbf{k}$-dependent terms relative to $q^2/2\mu_{\chi N}$ is maximized when $\mbf{q}\cdot\mbf{k} = -qk$. We see that these terms are small compared to $q^2/2\mu_{\chi N}$ when
\begin{equation}
    \frac{m_\text{N}}{\mu_{\chi N}}\gg\frac{k}{q}\left(2 + \frac{k}{q}\right).
\end{equation}
For light dark matter with $m_\chi \ll m_N$ the left hand side is bounded from below by $m_\text{N}/m_\chi$, and the soft limit corresponds to $(k/q)^2 \ll m_\text{N}/m_\chi$.
For dark matter masses $\mathcal{O}$(10--100) MeV typically probed by the Migdal effect, 
this is $k \ll 10q \lesssim \sqrt{m_\text{N}/m_\chi}q$, 
assuming a nucleus with mass $m_N\gtrsim 10$ GeV typical of detector materials. 
In this DM mass range typical momentum transfers are of the order $m_\chi v\gtrsim 10$ keV, suggesting that the soft limit would be valid roughly up to $k\sim 100$ keV. 

On the other hand, the ELF in Eq.~\eqref{eq:6D} decays fast for large values of $k$, and the primary contribution to the integral comes from the kinematic region, where $k\lesssim10$ keV. Combining this with the kinematic considerations from above, we see that the soft limit is valid across the kinematic region where the integrals receive meaningful contributions. We have further checked the validity of this approximation by comparing it with the exact expression in Eq.~\eqref{eq:6D}. In this way, we verified that in the mass range 10 MeV $\lesssim m_\chi\lesssim$ 1 GeV the error from these approximations is negligible compared to the error that arises from the numerical approximation of the integrals.

The angular structure of the integrand in Eq.~\eqref{eq:6D} can be resolved by noticing that 
the factor $(\mbf{\hat{q}} \cdot \mbf{\hat{k}})^2$ acts as a projection operator.  Writing it first as 
a linear combination of Legendre polynomials and then using the addition theorem of spherical harmonics, we obtain
\begin{equation}
    \label{eq:dot-to-sph-harm}
    (\unitv{q} \cdot \unitv{k})^2 = \frac13 P_0(\unitv{q} \cdot \unitv{k}) + \frac23 P_2(\unitv{q} \cdot \unitv{k}) = \frac{4\pi}{3} Y_{00}(\unitv{q}) Y_{00}(\unitv{k}) + \frac{8\pi}{15} \sum_{m=-2}^2 Y_{2m}(\unitv{q}) Y_{2m}(\unitv{k}).
\end{equation}
This will project the integrand in Eq.~\eqref{eq:6D} onto the direct sum of the $\ell=0$ and $\ell=2$ subspaces in the spherical harmonic basis. This allows us to write the spherical harmonic coefficients of the target response as
\begin{equation}
    T_{\ell m}(q,\omega_q,\omega) = \delta\left(\omega_q - \omega - \frac{q^2}{2m_N}\right)(\delta_{l0} + \delta_{l2})W_{\ell m}(\omega),
\end{equation}
where
\begin{equation}
    W_{\ell m}(\omega) = A_\ell \int \dr^3 \mbf{k}\, Y_{\ell m}(\unitv{k}) Z^2(k)  \mathcal{W}(\mbf{k}, \omega)=A_\ell\int dk k^2 Z(k)^2{\mathcal W}_{\ell m}(k), \label{eq:W_lm}
\end{equation}
with $A_0=4\pi/3$, $A_2 = 8\pi/15$.

Notice that apart from the first delta-function, this expression is now independent of $q$. Because of this, we may define the additional set of functions
\begin{equation}
    F_{\ell m}(\omega,v_\text{lab}(t)) = \int \dr^3 \mbf{q}\, q\hat{f}(s + \mbf v_\text{lab}\cdot\unitv{q}, \unitv{q})Y_{\ell m}(\unitv{q})=\int dq q^3 \hat{f}_{\ell m}(s, v_\text{lab}), \label{eq:F_lm}
\end{equation}
such that we can reduce the event rate, via equation \eqref{eq:gen-rate-harmonic}, to
\begin{equation}
    \label{eq:MigdalRateSeparated}
    \frac{\dr R_{\rm M}}{\dr\omega} = \frac{A^2 \alpha \bar\sigma_n \rho_\chi}{8\pi^4  \mu_{\chi n}^2 m_N^3 m_\chi \omega^4}\sum_{l=0,2} \sum_{|m|\leq l}W_{\ell m}(\omega)\mathcal{R}^{(\ell)}_{mm'}(t)F_{\ell m'}(\omega,v_\text{lab}(t)).
\end{equation}

For a multi-atomic target material, the rate has to be calculated for each nucleus type separately. The total rate is then the weighted sum
\begin{equation}\begin{aligned}
\frac{\dr R_{\rm M}}{\dr\omega} &= \sum_N \frac{M_N}{M_T} \frac{\dr R_N}{\dr\omega} ,
\end{aligned}\end{equation}
where $M_N = n_N m_N$ is the total mass of atoms of type $N$ in the target and $M_T$ the mass of the entire target.

The properties of the detector material are embedded in $W_{\ell m}$, while $F_{\ell m}$ contains the information of the DM velocity distribution. Eq.~\eqref{eq:F_lm} can be compared with the corresponding result in electron scattering, 
Eq.~\eqref{eq:electron-rate-short-harmonic}. The sensitivity of the rate to the angular dependence of the ELF (or the DM velocity distribution) is much more strictly constrained: the only anisotropic contribution to the event rate is from the quadrupole term ($\ell=2)$. 
For an isotropic material, the quadrupole term does not contribute and the rate simplifies to
\begin{equation}
    \frac{\dr R_{\rm M}^{\rm{iso}}}{\dr\omega} = \frac{A^2 \alpha \bar\sigma_n \rho_\chi}{8\pi^4  \mu_{\chi n}^2 m_N^3 m_\chi \omega^4} W_{00}(\omega)F_{00}(\omega,v_\text{lab}).
\end{equation}

If the velocity distribution is isotropic, the Radon transform $\hat{f}(s,\unitv{q})$ depends on $\unitv{q}$ only via the product $\mbf v_\text{lab}(t)\cdot\unitv{q}$. It follows that in the chosen integration coordinates, where $\mbf v_\text{lab}(t)$ is in the $z$-direction, $F_{\ell m}$ vanishes for $m\neq0$. In case of the SHM, expressions for the nonzero components of $F_{\ell m}$ are outlined in appendix \ref{ap:q-int}.

Our key result, equation \eqref{eq:MigdalRateSeparated}, is extremely convenient because it cleanly separates the domains and effects that contribute to the event rate into different expressions. Namely, $W_{\ell m}(\omega)$ captures the material response, the matrix $\mathcal{R}^{(\ell)}_{mm'}(t)$ describes  the effects due to the detector's orientation (daily modulation), and $F_{\ell m}(\omega,v_\text{lab}(t))$ captures the dark matter velocity distribution and effects of Earth's motion (annual modulation).

\section{Directional dependence of dark matter direct detection}
\label{section:results}

Anisotropies from the target response combined with Earth's motion through the DM halo lead to 
time-variation in the DM scattering event rates. These daily or annual modulation patterns give information about the directional dependence of the underlying phenomenon even if the detector design does not allow for reconstruction of the recoil direction for individual events. We will now determine the consequences of the formalism developed in the previous section, using the notation of Eq.~\eqref{eq:MigdalRateSeparated}. 

In practice, $W_{\ell m}(\omega)$ only needs to be evaluated once for a given material and can then be reused for different times, and for different parameterizations of the velocity distribution. Furthermore, the time variation of $F_{\ell m}(\omega,v_\text{lab}(t))$ is, to first order, determined by Earth's orbital motion with only a negligible contribution from motion of the lab around Earth's axis of rotation. It can therefore be sampled on a sparse time grid to capture the annual modulation, while the matrix $\mathcal{R}^2_{mm'}(t)$, which is cheap to evaluate, can be sampled on a denser grid to capture the daily modulation.

An implication of Eq.~\eqref{eq:MigdalRateSeparated} is that 
the expected magnitude of the daily modulation in the Migdal rate
can be estimated from the ratio $R_W=\|\mbf{W}_2(\omega)\|/|W_{00}(\omega)|$, where
\begin{equation}
    \|\mbf{W}_2(\omega)\|^2 = \sum_{|m|\leq 2}|W_{2m}(\omega)|^2.
    \label{eq:rateestimator}
\end{equation}
A material can therefore be identified as a potentially suitable candidate for observing daily modulation without a need for evaluation of the event rate if this ratio is large enough. However, note that this is not a definitive determination, because the actual modulation of the event rate could possibly still be suppressed or amplified depending on the values of $F_{2m}(\omega, \mbf v_\text{lab}(t))$ relative to $F_{00}(\omega, \mbf v_\text{lab}(t))$. However, a ratio $R_W\ll 1$ indicates that daily modulation is unlikely to be observable in the material at hand. 

It is also notable that the coefficients $W_{\ell m}(\omega)$ are related to the spherical harmonic moments of the ELF. The fact that the event rate only depends on the $\ell = 0, 2$ moments therefore implies that the effects of the dipole moment ($\ell = 1$), and moments higher than the quadrupole, on the daily modulation are heavily suppressed in the Migdal rate. For observation of daily modulation of the Migdal rate, one should therefore look for a detector material whose symmetries would give rise to a large quadrupole ($\ell = 2$) contribution to the dielectric function and, by extension, to the ELF.

\subsection{General features}

As discussed above, the daily modulation of the Migdal rate is generated, to a good approximation, solely by the quadrupole component of the ELF. In contrast, in electron scattering all low-order harmonics contribute. To illustrate the consequences of this behavior, we use a toy model of the ELF with particular angular features. We consider separately cases with anisotropy given by the dipole or the quadrupole term,
\begin{equation}\begin{aligned}
&\mathcal W_1 = 
\exp\left[-\left(\frac{k}{10\text{keV}}\right)^2\right]  \left( 1 + \sqrt\frac{4\pi}{3} Y_{1,1}(\theta,\phi) \right),\qquad 
\mathcal W_2 = 
\exp\left[-\left(\frac{k}{10\text{keV}}\right)^2\right]  \left( 1 + \sqrt\frac{16\pi}{15} Y_{2,1}(\theta,\phi) \right) .
\label{eq:toyW}
\end{aligned}\end{equation}
To further illustrate the general features, we consider another parametric model, which receives contributions from all multipoles in its spherical expansion,
\be 
\mathcal{W}_3 = \exp\left[-\left(\frac{k}{10\text{keV}}\right)^2\right] \left( 1+ \sin(\theta \phi) \right).
\label{eq:toyWgeneric}
\ee 
Here we also allow for the separability of the $k$-dependence and angular dependence to make these toy models maximally anisotropic in the sense that they vanish for certain directions for every $k$. This will maximize the amplitude of the daily modulation.

We select Gran Sasso as the location of our hypothetical detector, oriented in a right-handed coordinate system such that the $z$-axis points towards the zenith, and the $x$-axis points west. The laboratory velocities $\mbf v_\text{lab}$ were calculated for each hour on August 20, 2027. 
From these velocities the event rates were computed analytically using Eqs. \eqref{eq:electron-rate-short} and \eqref{eq:MigdalRateSeparated}.

\begin{figure}
\centering
\includegraphics[width=0.49\textwidth]{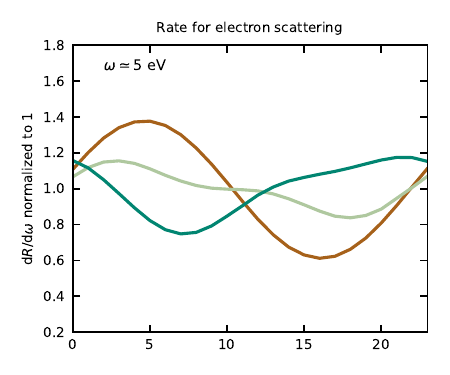}
\includegraphics[width=0.49\textwidth]{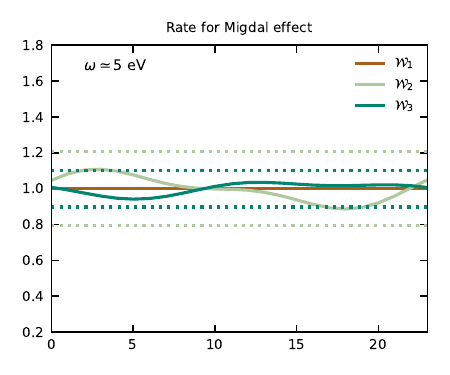}
\includegraphics[width=0.49\textwidth]{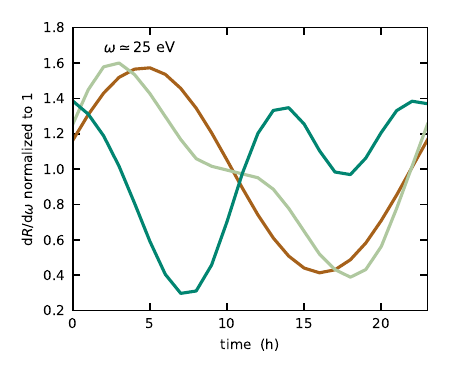}
\includegraphics[width=0.49\textwidth]{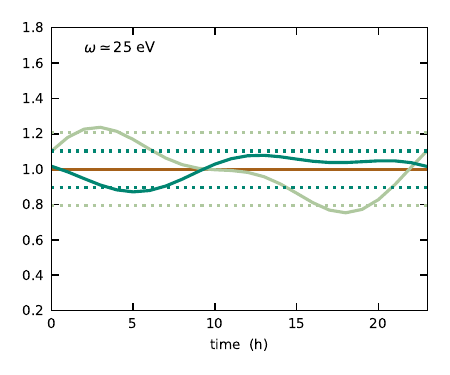}
\caption{Daily modulation of the event rate for parametric model ELFs in Eq.~\eqref{eq:toyW} and \eqref{eq:toyWgeneric}, assuming the SHM velocity distribution and DM mass of 10\,MeV. The left panel shows the result for electron recoils and the right panel shows the result for the Migdal effect. The horizontal lines in the right panel correspond to $1 \pm R_W$. The top row is for $\omega \simeq 5 \rm{\,eV}$ and the bottom row is for $\omega \simeq 25 \rm{\, eV}$.} 
\label{fig:toy}
\end{figure}

The results for the daily modulation arising from these parametric models are shown in Fig.~\ref{fig:toy}. The left panel of the Figure shows the event rate for the electron recoils and the right panel for the Migdal effect, assuming the standard halo model for the DM velocity distribution. We see that the full calculation gives results compatible with our main results given in Eqs.~\eqref{eq:electron-rate-short-harmonic} and~\eqref{eq:MigdalRateSeparated}. The angular anisotropy due to the quadrupole term ($\ell=2$) is similar in both cases, while the dipole term ($\ell=1$) is only relevant for electron scattering.

The rows in Fig.~\ref{fig:toy} show the daily modulation for different $\omega$. As can be seen, the modulation increases with the energy. In the Migdal case, this can be understood through the velocity distribution integral: for the SHM, the tail is more anisotropic. This makes $F_{20}(\omega)/F_{00}(\omega)$ an increasing function that indirectly scales $R_W$ to give the order of the modulation amplitude. Since for the SHM, $F_{20}$ and $F_{00}$ are of the same order, $R_W$ gives an estimate on the modulation amplitude for the Migdal rate.
However, the rate decreases with the energy $\omega$. Therefore, even though higher values of $\omega$ display larger relative modulation amplitude, the absolute rate is lower and the prospects for observing the daily modulation must be balanced with respect to the relative amplitude and the absolute rate.

\subsection{Silicon and gallium arsenide}

We calculated the electron scattering and Migdal effect rates for silicon (Si) and gallium arsenide (GaAs) detectors, which are widely adopted target materials for leading and next-generation sub-GeV dark matter experiments such as SuperCDMS \cite{SuperCDMS2}, SENSEI \cite{SENSEI}, TESSERACT \cite{TESSERACT}, DAMIC-M \cite{DAMIC-M} and DAREDEVIL \cite{DAREDEVIL,DAREDEVIL2}. 
An accurate evaluation of such rates requires a precise description of the target materials' ELFs, which we computed from first principles following the procedure described in Sec.~\ref{subsection:DFT}.

The DFT calculations were performed using the PBE functional \cite{PBE} and a $20\times20\times20$ $\mathbf{k}$-point grid. A scissor correction was applied to set the energy band gap to the experimental value: $E_{\rm gap}=1.12$ eV for Si \cite{Sigap} and $E_{\rm gap}=1.42$ eV for GaAs \cite{GaAsgap}.
The dielectric response was calculated within the adiabatic local density approximation (ALDA) \cite{ALDA}, which approximates the dynamic exchange-correlation kernel $f_{xc}$ using the static, local limit (the RPA approximation neglects instead this contribution). The calculation employed an energy broadening of $0.001$ eV and included 70 unoccupied bands to ensure the response's convergence. To optimize computational cost, the response was evaluated exclusively on the irreducible points of a $20\times20\times20$ $\mathbf{k}$-point grid, and subsequently reconstructed over the full BZ using the symmetry transformations previously described. A momentum cutoff of $q_{\text{cut}} \approx 20 \text{ keV}$ was applied by setting a kinetic energy cutoff  $E_{\text{cut}} =\frac{q_{\text{cut}}^2}{2m_e}=400 \text{ eV}$. Local field effects were included by using Eq. \eqref{eq:LFEs}. 
Finally, for the evaluation of the Migdal effect rate, the momentum dependent effective ion charges $Z(k)$ were extrapolated from the experimental data of Ref. \cite{ZionSi}. Tab. \ref{tab:dft_parameters} collects the main parameters associated with the evaluation of the ELF.

\begin{table}[htbp]
\setlength{\tabcolsep}{7pt}
\centering
\label{tab:dft_parameters}
\begin{tabular}{lcc}
\toprule
Material & Si & GaAs \\\midrule
$\mathbf{k}$-point grid & $20 \times 20 \times 20$ & $20 \times 20 \times 20$ \\
xc-functional & PBE & PBE \\
$E_{\text{cut}}$ (eV) & 400 & 400 \\
Empty states & 70 & 70 \\
xc-kernel ($f_{\text{xc}}$) & ALDA & ALDA \\
$E_{\text{gap}}$ (eV) & 1.12 & 1.42 \\
\bottomrule
\end{tabular}
\caption{Key electronic structure and convergence parameters used to evaluate the energy loss function of Si and GaAs.}
\end{table}

The rates were evaluated using Eqs.~\eqref{eq:electron-rate-short} and \eqref{eq:W_lm}--\eqref{eq:MigdalRateSeparated} for the SHM velocity distribution. Given that the ELF from the DFT simulations comes sampled on a momentum grid, we performed integrals involving the ELF using a simplex integration method over a mesh produced via Delaunay triangulation. In case of the Migdal effect, the radial $q$-integrals in the evaluation of $F_{\ell m}(\omega,v_\text{lab})$ were performed via conventional adaptive quadrature.

We use a low DM mass $m_\chi = 10 \text{\,MeV}$, as the daily modulation effect is expected to be the greatest for low-energy scattering events. We chose the values $\bar\sigma_n = 10^{-36} \text{\,cm}^2$ and $\bar\sigma_e = 10^{-40} \text{\,cm}^2$ for our reference cross sections, as representative of the parameter values allowed by existing limits. The precise value of the cross section is not relevant for this analysis, since it simply scales the overall rate and does not affect the shape of the energy spectrum or the daily modulation.

\begin{figure}[b!]
    \centering
    \includegraphics[width=0.49\linewidth]{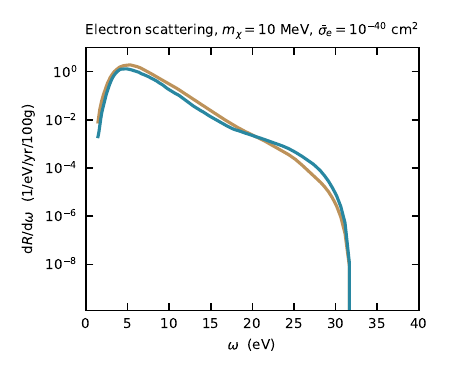}
     \includegraphics[width=0.49\linewidth]{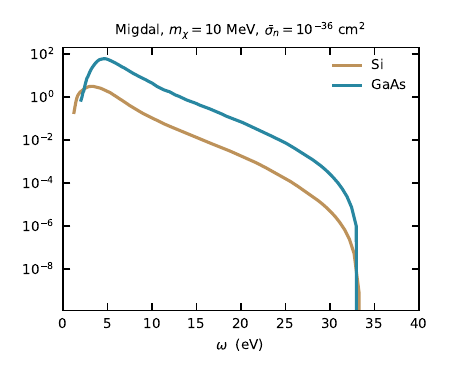}
    \caption{The electron scattering (left) and Migdal (right) event rates in silicon and gallium arsenide for the SHM velocity distribution and a DM mass of 10 MeV.}
    \label{fig:rates}
\end{figure}

The resulting event rates for Si and GaAs are shown in Fig.~\ref{fig:rates}.
The electron scattering rates are found compatible with previous results within the level of uncertainty expected due to the computational methodology for the ELF \cite{Knapen,DarkELF,Griffin_PRD}. Further refinements could include higher-energy transitions from core states or extensions to larger momentum transfers \cite{Griffin_PRD}. However, core transitions are kinematically forbidden for lower DM masses and, lacking momentum dispersion, yield purely isotropic contributions. Likewise, extending the high-$q$ cutoff is computationally prohibitive without reverting to isotropic approximations, such as evaluating the ELF over a line in momentum space, or neglecting LFEs. Consequently, while these high-energy and high-momentum contributions can quantitatively modify the total rate, increasing it for massive mediators and heavier DM masses, they will not produce any relevant difference in the overall rate anisotropy.

We note that the Migdal rate in Si matches values found in the literature \cite{Berghaus} rather well. We can see that the electron recoil rates for both materials are relatively similar, although the rate in GaAs decreases slower at energies above 15 eV. The Migdal rates differ between the materials mainly due to the different effective ion charges $Z(k)$. The maximum contribution to both event rates comes relatively close to the minimum energy at around $\omega\simeq 3  \rm{\, eV}$ to $5 \rm{\, eV}$.

\begin{figure}
    \centering
    \includegraphics[width=0.49\linewidth]{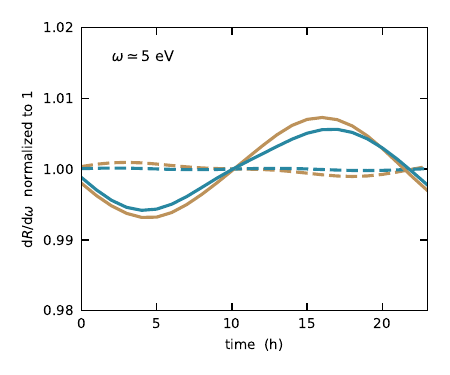}
    \includegraphics[width=0.49\linewidth]{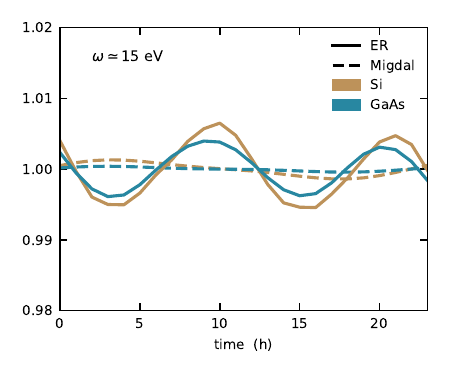}
    \caption{Electron scattering (solid) and Migdal (dashed) daily modulation in silicon (orange) and gallium arsenide (blue) at energies $\omega \simeq 5 \rm{\,eV}$ (left) and $\omega \simeq 25 \rm{\, eV}$ (right) on August 20, 2027, assuming the SHM velocity distribution and a DM mass of 10\,MeV.}
    \label{fig:day}
\end{figure}

The daily modulation in Si and GaAs is shown in Fig.~\ref{fig:day}. We first note that the overall scale of the modulation is relatively small for both materials, on the order of 1\% in the electron recoil case. This is reflective of the fact that the ELFs for these materials are nearly isotropic. The modulation of the electron rate at 5 eV appears roughly like a clean sine wave with a period of one sidereal day, suggesting that it originates primarily from the dipole component. This is consistent with the degree of suppression of the Migdal rate.

\section{Conclusions}
\label{section:checkout}

We have developed a general framework for evaluating dark matter induced electronic excitations in solids while retaining the full anisotropic structure of the material response. In contrast to conventional treatments based on isotropically averaged dielectric functions, our formulation preserves the directional dependence encoded in the energy loss function (ELF), allowing a direct connection between crystal symmetry, electronic structure, and observable dark matter signatures.

Using a spherical harmonic decomposition, we showed that the angular structure of the target response maps directly onto the temporal modulation pattern of the event rate. In particular, for the Migdal effect the daily modulation is controlled predominantly by the quadrupolar component of the ELF, whereas direct DM--electron scattering can retain sensitivity to a broader set of angular harmonics. This decomposition not only provides a practical route for event-rate calculations but also offers physical insight into which material properties control directional sensitivity.

Our results suggest a conceptual shift in the study of Migdal processes in solids. Historically, the Migdal effect has been viewed largely as a mechanism that enhances rates at low dark matter masses. In solids, however, the Migdal effect  becomes sensitive not only to electronic excitation energies but also to the geometry and symmetry of wave functions throughout momentum space. 
The weak modulation observed in conventional semiconductors such as silicon and gallium arsenide originate from the relatively isotropic electronic responses, and thus point  toward the search of new class of candidate detector materials where stronger directional signatures may emerge.
The present formalism and computational approach can thus be used to explore particularly promising systems with intrinsically anisotropic dielectric properties or strongly directional low-energy excitations. 
Particularly promising candidates include layered and low-symmetry semiconductors, van der Waals heterostructures, topological insulators, and materials near electronic instabilities, where strong momentum-space anisotropy coexists with experimentally favorable detector properties. 
Similarly, correlated materials and altermagnets may exhibit unconventional angular signatures associated with orbital and spin textures.
In addition, given the lack of the coherent scattering scaling that typically favors heavy element targets for nuclear recoils, low mass materials, specifically hydrogen-based materials like hydrides, represents an interesting research line \cite{H_Migdal}. 
Quantities such as quadrupole moments of the ELF or symmetry-resolved dielectric descriptors may serve as screening criteria in high-throughput computational searches. Combined with modern first-principles methods and materials databases, this could enable systematic identification of optimal targets for directional dark matter experiments.

Several theoretical directions may naturally follow from the present work. First, extending the formalism beyond the isotropic Standard Halo Model may reveal stronger directional signatures. Second, the present treatment can be generalized to velocity-dependent interactions and broader effective field theory operators. Third, a fully momentum-resolved description of the Migdal process beyond the soft approximation may become important in materials with sharp reciprocal-space features. Furthermore, inclusion of electron-phonon coupling, excitonic effects, and strong electronic correlations may substantially alter the dielectric response and therefore the predicted signal.
We expect that future detectors may exploit engineered anisotropic responses transforming directional sensitivity into a promising strategy for identifying dark matter.

\begin{acknowledgments}
This work has been supported by the Research Council of Finland (grants\# 342777, 371542). AA thanks Wihuri foundation for financial support. 
\end{acknowledgments}

\appendix

\section{Momentum transfer integrals}\label{ap:q-int}

In the simplest case, we have the standard halo model as the velocity distribution. In the Migdal rate, two of the three dimensions in the nucleus momentum integral
\begin{align}
    F_{\ell m}(\omega,\mbf v_\text{lab}(t)) &= \int \dr^3 \mbf{q}\, q\hat{f}_\text{SHM}(s, \unitv{q})Y_{\ell m}(\unitv{q}) 
\end{align}
can be solved analytically. First, we note that with an appropriate choice of integration coordinates, such that the $z$-axis is along $\mbf v_\text{lab}$, the only part the integral which depends on the azimuthal angle $\phi$ is
\begin{equation}\begin{aligned}
    \int_0^{2\pi}\dr\phi\, Y_{\ell m}(\theta,\phi).
\end{aligned}\end{equation}
By definition of the spherical harmonics, we can see that this vanishes for $m\neq 0$. Therefore the only components that remain are
\begin{equation}
    F_{\ell 0} = 2\pi \int_0^\infty \dr q \, q^3 \int_{-1}^1\dr\cos\theta \left( \exp\left[-\frac{(s+ v_\text{lab} \cos\theta)^2}{v_0^2}\right] - \exp\left[-\frac{v_{\rm esc}^2}{v_0^2}\right] \right) \Theta(v_{\rm esc}-|s+ v_\text{lab} \cos\theta|) Y_{\ell 0}(\theta,\phi) .
\end{equation}

The step function $\Theta$ imposes some nontrivial limits on the integral. Denoting $x=\cos\theta$, we may solve from the limits from the equations $|x|\leq 1$ and $|s+v_\text{lab}x|\leq v_\text{esc}$. We find
\begin{align}
    x_\text{min}&=\max\left\{-1,-\frac{v_\text{esc}+s}{v_\text{lab}}\right\},\\
    x_\text{max}&=\min\left\{1,\frac{v_\text{esc}-s}{v_\text{lab}}\right\}.
\end{align}
In practice, since $v_\text{esc}>v_\text{lab}$, the lower limit typically reduces to $x_\text{min}=-1$. These limits imply that
\begin{equation}
    -v_\text{lab}\leq v_\text{esc}-s.
    \label{eq:slimit}
\end{equation}
Given the approximation
\begin{equation}
    s\approx \frac{q}{2\mu}+\frac{\omega}{q} ,
\end{equation}
we can use \eqref{eq:slimit} to solve the limits on the $q$-integral, and we find that
\begin{align}
    q_\text{min}&=\mu(v_\text{lab}+v_{\rm esc})-\sqrt{\mu^2(v_\text{lab}+v_{\rm esc})^2-2\mu\omega},\\
    q_\text{max}&=\mu(v_\text{lab}+v_{\rm esc})+\sqrt{\mu^2(v_\text{lab}+v_{\rm esc})^2-2\mu\omega}.
\end{align}
Notice that this also implies that $\mu^2(v_\text{lab}+v_{\rm esc})^2-2\mu\omega \geq 0$, leading to a limit
\begin{equation}
    \omega_\text{max}=\frac{1}{2}\mu(v_\text{lab}+v_\text{esc})^2
\end{equation}
above which the event rate vanishes.

Given that
\begin{equation}
    Y_{\ell 0}(\theta,\phi)=\sqrt{\frac{2\ell+1}{4\pi}}P_\ell(\cos\theta),
\end{equation}
where $P_\ell(x)$ are the Legendre polynomials, the angular integral reduces to
\begin{equation}\begin{aligned}
    I_\ell(q)=\sqrt{\frac{2\ell+1}{4\pi}}\int_{-1}^{x_\text{max}} \dr x \left( \exp\left[-\frac{(s+ v_\text{lab} x)^2}{v_0^2}\right] - \exp\left[-\frac{v_{\rm esc}^2}{v_0^2}\right] \right) P_\ell(x)
\end{aligned}\end{equation}
has to be written separately for the cases $\ell=0,2$, with $P_0(x)=1$ and $P_2(x)=\tfrac{1}{2}(3x^2-1)$ we have
\begin{equation}\label{eq:I0}
    I_0(q)=
    \begin{cases}
        \displaystyle\frac{\sqrt\pi v_0}{2 v_\text{lab}} \left[ \mathrm{erf}\left(\frac{s+v_\text{lab}}{v_0}\right) - \mathrm{erf}\left(\frac{s-v_\text{lab}}{v_0}\right)\right] - 2\exp\left(\frac{-v_{\rm esc}^2}{v_0^2}\right),& x_\text{max}=1\\[0.5em]
        \displaystyle\frac{\sqrt\pi v_0}{2 v_\text{lab}} \left[\mathrm{erf}\left(\frac{v_{\rm esc}}{v_0}\right)- \mathrm{erf}\left(\frac{s-v_\text{lab}}{v_0}\right)\right] - \exp\left(-\frac{v_{\rm esc}^2}{v_0^2}\right) \left(\frac{v_{\rm esc}-s}{v_\text{lab}}+1\right),& x_\text{max}=\frac{v_\text{esc}-s}{v_\text{lab}}\\
    \end{cases}
\end{equation}
and
\begin{equation}
    I_2(q) = \frac{v_0^3}{4v_\text{lab}^3}
    \begin{cases}
        \begin{aligned}
            &\left\{\sqrt\pi \left(6\frac{s^2}{v_0^2} -2\frac{v_\text{lab}^2}{v_0^2} +3\right) \left[\erf\left(\frac{s+v_\text{lab}}{v_0}\right) - \erf\left(\frac{s-v_\text{lab}}{v_0}\right) \right]\right. \\
            &\quad{}+\left.6\frac{s-v_\text{lab}}{v_0} \exp(\frac{-(s+v_\text{lab})^2}{v_0^2}) -6\frac{s+v_\text{lab}}{v_0} \exp(\frac{-(s-v_\text{lab})^2}{v_0^2}) \right\},\\[1em]
        \end{aligned}& x_\text{max}=1\\
        \begin{aligned}
            &\left\{\sqrt\pi (6\frac{s^2}{v_0^2} -2\frac{v_\text{lab}^2}{v_0^2} +3) \left[\erf\left(\frac{v_{\rm esc}}{v_0}\right)-\erf\left(\frac{s-v_\text{lab}}{v_0}\right)\right]\right. \\
            &\quad{}+\left[ 4\frac{v_\text{lab}^2(v_{\rm esc}-s)-(v_{\rm esc}-s)^3}{v_0^3} +6\frac{2s-v_{\rm esc}}{v_0} \right] \exp(\frac{-v_{\rm esc}^2}{v_0^2}) \\
            &\quad{}-\left.6\frac{s+v_\text{lab}}{v_0} \exp(\frac{-(s-v_\text{lab})^2}{v_0^2}) \right\},
        \end{aligned}&\displaystyle x_\text{max}=\frac{v_\text{esc}-s}{v_\text{lab}}\\
    \end{cases}
\end{equation}
The radial integrals
\begin{equation}
    F_{\ell 0}(\omega,\mbf v_\text{lab}(t)) = \int_{q_\text{min}}^{q_\text{max}}\dr q\, q^3I_\ell(q)
\end{equation}
are left to be evaluated numerically.

\bibliography{Bibliography}
\end{document}